\documentclass[11pt,a4paper]{article}

\usepackage[utf8]{inputenc}
\usepackage[T1]{fontenc}
\usepackage{lmodern}
\usepackage[margin=2.7cm]{geometry}
\usepackage{amsmath,amssymb,amsthm}
\usepackage{mathtools}
\usepackage{booktabs}
\usepackage{caption}
\usepackage{subcaption}
\usepackage{enumitem}
\usepackage{xcolor}
\usepackage{tikz}
\usetikzlibrary{arrows.meta,positioning,calc,fit,backgrounds}
\usepackage{tikz-cd}
\usepackage[square,comma,sort]{natbib}
\usepackage[colorlinks=true,linkcolor=blue!50!black,citecolor=green!35!black,urlcolor=blue!50!black]{hyperref}
\usepackage{url}

\newcommand{\N}{\mathbb{N}}
\newcommand{\R}{\mathbb{R}}
\newcommand{\Set}{\mathsf{Set}}
\newcommand{\Petri}{\mathsf{Petri}}
\newcommand{\CMC}{\mathsf{CMC}}

\newcommand{\OpenPetri}{\mathsf{Open}(\mathsf{Petri})}
\newcommand{\DOpenPetri}{\mathbb{O}\mathbf{pen}(\mathsf{Petri})}
\newcommand{\DOpenCMC}{\mathbb{O}\mathbf{pen}(\mathsf{CMC})}
\newcommand{\RxNet}{\mathsf{RxNet}}
\newcommand{\Dynam}{\mathsf{Dynam}}
\DeclareRobustCommand{\pto}{\mathrel{\ooalign{\hfil$\mapstochar\mkern5mu$\hfil\cr$\to$\cr}}}
\newcommand{\Nset}[1]{\N[#1]}
\newcommand{\id}{\mathrm{id}}
\newcommand{\T}[2]{T_{#1,#2}}

\theoremstyle{definition}
\newtheorem{definition}{Definition}
\newtheorem{example}[definition]{Example}
\theoremstyle{remark}
\newtheorem{remark}[definition]{Remark}

\definecolor{placeblue}{RGB}{31,78,140}
\definecolor{transgreen}{RGB}{20,130,60}
\definecolor{boxcol}{RGB}{40,90,105}
\tikzset{
  place/.style={circle, draw=placeblue, very thick, minimum size=8.5mm, inner sep=0pt, font=\small},
  vtrans/.style={rectangle, draw=transgreen, very thick, minimum width=3mm, minimum height=9mm, inner sep=0pt},
  htrans/.style={rectangle, draw=transgreen, very thick, minimum width=10mm, minimum height=3.2mm, inner sep=0pt, font=\scriptsize},
  dtrans/.style={rectangle, draw=transgreen, very thick, dashed, minimum width=14mm, minimum height=3.6mm, inner sep=1pt, font=\scriptsize},
  degr/.style={rectangle, draw=transgreen, thick, minimum width=2.2mm, minimum height=5mm, inner sep=0pt},
  arc/.style={-{Stealth[length=2.2mm]}, thick},
  rarc/.style={{Stealth[length=2.2mm]}-{Stealth[length=2.2mm]}, thick},
  leg/.style={-{Stealth[length=1.8mm]}, densely dashed, gray!70!black},
  wt/.style={font=\scriptsize, inner sep=1pt},
  bnd/.style={rectangle, draw=boxcol, very thick, minimum width=6mm, inner sep=0pt},
  pt/.style={circle, fill=black, inner sep=0pt, minimum size=3pt},
  lab/.style={font=\small}
}
\newcommand{\tok}{\textcolor{red}{$\bullet$}}

\title{An Open Petri Net Implementation\\ of Gene Regulatory Networks}
\author{Yanying Wu\thanks{\texttt{yanying.wu@cncb.ox.ac.uk}}\\[4pt]
\small Centre for Neural Circuits and Behaviour, University of Oxford, UK\\
\small Department of Physiology, Anatomy and Genetics, University of Oxford, UK}
\date{18 November 2026}

\begin{document}
\maketitle

\begin{abstract}
Gene regulatory networks (GRNs) are a central object of systems biology, and many formalisms have been used to model them, among them Petri nets. Ordinary Petri nets, however, provide no built-in way to assemble a large network from smaller, separately specified pieces. Open Petri nets, in the sense of Baez and Master, supply exactly this: an open Petri net is a Petri net with designated boundary places, and open Petri nets are composed by gluing boundaries together (a pushout), giving a symmetric monoidal double category $\DOpenPetri$. In this paper we (i) state precisely the standard encoding of activation, inhibition and degradation as Petri net transitions; (ii) distinguish the classical ``Petri nets are monoids'' view of Meseguer and Montanari, which describes how \emph{processes inside one net} compose, from the open Petri net view, which describes how \emph{nets} compose; (iii) show, with a worked example, how a GRN can be cut into open Petri net modules along shared genes and recovered exactly by composition, together with simple rules that make such a decomposition correct; and (iv) use the compositional rate-equation semantics of Baez and Pollard to show that the mass-action dynamics of the composite network are obtained by gluing the dynamics of its modules, linking Petri net models of GRNs to their ordinary differential equation counterparts. We close by discussing what the reachability semantics does and does not give for GRN modules, and the limitations of the approach.
\end{abstract}

\section{Introduction}
The human genome contains about 20,000 protein-coding genes, whose expression changes across tissues and throughout life, from development to ageing. In any given cell, a large number of gene products interact with one another, forming a complicated network that supports the processes of life. Such a network is called a \emph{gene regulatory network} (GRN), and it has been studied for decades in genomics and systems biology.

Many computational approaches have been developed to model GRNs, including information-theoretic models, Boolean networks, differential equation models, Bayesian networks and Petri nets \citep{deJong2002,Delgado2019,Steggles2007}. Petri nets are attractive for this purpose: they are designed for concurrent dynamical systems, they have a large user community, and they come with a well-developed theory \citep{Reisig1985,Murata1989,Steggles2007,BordonMraz2012}.

A standard Petri net specifies the structure and behaviour of one particular system. It is, however, not convenient to compose larger nets from smaller ones, since ordinary Petri nets lack a notion of interface along which nets can be connected, and lack mechanisms for abstraction \citep{ErmelMartini1996}. Large systems such as GRNs would benefit from being specified as modules that interact only through well-defined interfaces, so that each module can be built, checked and reused on its own. Category theory provides a natural language for this, and a line of work going back to \citet{MeseguerMontanari1990} has studied Petri nets categorically \citep{ErmelMartini1996,Baldan2005,Rathke2014,Baldan2015,BaezPollard2017,BaezMaster2020}. Within this line, the \emph{open Petri nets} of \citet{BaezMaster2020} make composition of nets along shared places precise.

Standard Petri nets and their variants have been used to model GRNs for some time \citep{Chaouiya2006,Chaouiya2007,Chaouiya2008,Steggles2007,Ruths2008,BordonMraz2012,LiuHeiner2014,LiuHeinerGilbert2019,Hamed2018}. In a parallel development, the \emph{decorated cospan} formalism of \citet{Fong2015}, of which open Petri nets are a close relative, has been used to give compositional frameworks for electrical circuits \citep{BaezFong2018}, Markov processes \citep{BaezFongPollard2016} and chemical reaction networks \citep{BaezPollard2017}; the last of these is formulated precisely in terms of open Petri nets with rates. To our knowledge, open Petri nets have not yet been applied to GRNs. In this work we explore what such an application involves and what it offers, with the aim of bringing compositionality to GRN modelling.

\paragraph{Contributions and changes from version 1.} This version corrects several errors in the categorical part of version 1 and fills the main gap of that version, which presented the composition of two abstract open Petri nets but never applied it to a GRN. Concretely:
\begin{itemize}[nosep]
\item Section~\ref{sec:pn-grn} states the Petri net encoding of regulatory interactions precisely, adds degradation, and records its limitations.
\item Section~\ref{sec:monoids} explains what the ``Petri nets are monoids'' view does: it gives an algebra of \emph{processes inside a fixed net}, with correct examples of sequential and parallel composition.
\item Section~\ref{sec:open} presents open Petri nets, their composition by pushout and the double category $\DOpenPetri$, with corrected composition formulas, and summarises their operational, reachability and rate-equation semantics.
\item Section~\ref{sec:grn-open} applies this to GRNs: a GRN is decomposed into two open Petri net modules along shared genes and recovered exactly by composition, and its mass-action rate equation is obtained by gluing those of the modules.
\end{itemize}

\section{Gene regulatory networks}\label{sec:grn}
To support normal cellular function, genes interact with each other to carry out many crucial molecular processes. These interactions usually involve regulation of gene expression: if a change in the expression of gene $X$ can induce a change in the expression of gene $Y$, we say that $X$ regulates $Y$. Regulation can increase (activation, up-regulation) or decrease (inhibition, down-regulation) the expression of the target. Genes together with their regulatory relationships form the gene regulatory network \citep{Chai2014}.

GRNs can be inferred from gene expression data using approaches such as ordinary differential equations, Boolean networks and Bayesian networks, each with its own strengths and weaknesses \citep{deJong2002,Chai2014}. \emph{Ordinary differential equation} (ODE) models use continuous variables for the concentrations of mRNAs or proteins and describe how these concentrations change in time. They are well suited to quantitative analysis of nonlinear behaviour, but they are comparatively complex and their kinetic parameters are often unknown in practice. \emph{Boolean networks} represent each gene as either on or off and each regulatory rule as a Boolean function of the states of its regulators. They are simple, but they cannot capture many quantitative details. \emph{Probabilistic Boolean networks} extend Boolean networks by allowing several candidate Boolean functions per gene, one of which is chosen at random at each update \citep{Chai2014}.

\emph{Bayesian networks} combine probability and graph theory. A Bayesian network represents a GRN as a directed acyclic graph $G=(V,A)$, where the nodes $V=\{x_1,\dots,x_n\}$ are random variables for gene expression levels and the edges $A$ encode conditional dependencies. Both the graph structure and the conditional probability parameters are learned from expression data, typically by score-based search or constraint-based methods, and prior biological knowledge can be included through priors \citep{Chai2014}. Bayesian networks have been used successfully to improve the quality of inferred GRNs \citep{Chai2014}. Because the graph must be acyclic, static Bayesian networks cannot represent feedback loops, which are common in GRNs; dynamic Bayesian networks, which unroll the network over discrete time steps, remove this restriction \citep{deJong2002,Chai2014}.

\section{Petri net modelling of gene regulatory networks}\label{sec:pn-grn}
Petri nets offer a modelling style complementary to the methods above. They were introduced by C.~A.~Petri to model distributed and concurrent systems \citep{Petri1962,Petri1966}. Petri nets combine a well-defined mathematical theory, which allows precise analysis of system behaviour, with a graphical representation, which gives an intuitive picture of the structure and state changes of the system. They have been applied to many kinds of dynamical systems, including computer networks, communication systems and logistics \citep{LiuHeiner2014}.

\subsection{Petri nets and the firing rule}
A Petri net is a bipartite directed graph with two kinds of vertices, \emph{places} and \emph{transitions}; arcs connect places to transitions and transitions to places, but never two places or two transitions. Each arc carries a positive integer \emph{weight} (1 if not shown). Figure~\ref{fig:simple} shows a Petri net with two places $P_1,P_2$ and one transition $T_1$.

A \emph{marking} (or state) assigns a natural number of \emph{tokens} to each place; in Figure~\ref{fig:simple}, $P_1$ holds one token. A transition is \emph{enabled} when each of its input places holds at least as many tokens as the weight of the arc from that place. An enabled transition may \emph{fire}: it removes from each input place as many tokens as the weight of the corresponding arc and adds to each output place as many tokens as the weight of the corresponding arc, producing a new marking (Figure~\ref{fig:firing}). Firing need not conserve the total number of tokens \citep{Petri1966}.

Formally, we will use the following definition, which is the one used by \citet{MeseguerMontanari1990} and \citet{BaezMaster2020}. For a set $S$, let $\Nset{S}$ denote the set of finite formal linear combinations of elements of $S$ with natural number coefficients, i.e.\ the free commutative monoid on $S$ (equivalently, finite multisets over $S$).

\begin{definition}\label{def:petri}
A \emph{Petri net} consists of a set $S$ of places, a set $T$ of transitions, and two functions $s,t\colon T\to\Nset{S}$, the \emph{source} and \emph{target}. A \emph{marking} is an element $m\in\Nset{S}$. A transition $\tau$ is \emph{enabled} at $m$ if $m\ge s(\tau)$ coefficientwise, and then \emph{firing} $\tau$ produces the marking $m-s(\tau)+t(\tau)$.
\end{definition}

The coefficient of a place $p$ in $s(\tau)$ is the weight of the arc from $p$ to $\tau$ (0 if there is no arc), and the coefficient of $p$ in $t(\tau)$ is the weight of the arc from $\tau$ to $p$. For example, the transition in Figure~\ref{fig:firing} has $s(T_1)=P_1+2P_2$ and $t(T_1)=3P_3+P_4+2P_5$.

\begin{figure}[t]
\centering
\begin{tikzpicture}
  \node[place,label=right:$P_1$] (p1) at (0,0) {\tok};
  \node[htrans,label=right:{\small$T_1$}] (t1) at (0,-1.3) {};
  \node[place,label=right:$P_2$] (p2) at (0,-2.6) {};
  \draw[arc] (p1) -- (t1);
  \draw[arc] (t1) -- (p2);
\end{tikzpicture}
\caption{A simple Petri net with one token in place $P_1$.}\label{fig:simple}
\end{figure}

\begin{figure}[t]
\centering
\begin{tikzpicture}[scale=0.95]
  \begin{scope}
  \node[place,label=right:$P_1$] (p1) at (-0.9,0) {\tok};
  \node[place,label=right:$P_2$] (p2) at (0.9,0) {\tok\tok};
  \node[htrans,label=right:{\small$T_1$}] (t) at (0,-1.3) {};
  \node[place,label=below:$P_3$] (p3) at (-1.3,-2.7) {};
  \node[place,label=below:$P_4$] (p4) at (0,-2.7) {};
  \node[place,label=below:$P_5$] (p5) at (1.3,-2.7) {};
  \draw[arc] (p1) -- (t);
  \draw[arc] (p2) -- node[wt,right]{2} (t);
  \draw[arc] (t) -- node[wt,left]{3} (p3);
  \draw[arc] (t) -- (p4);
  \draw[arc] (t) -- node[wt,right]{2} (p5);
  \end{scope}
  \draw[arc] (2.3,-1.3) -- node[above,font=\small]{$T_1$ fires} (4.0,-1.3);
  \begin{scope}[xshift=6.3cm]
  \node[place,label=right:$P_1$] (q1) at (-0.9,0) {};
  \node[place,label=right:$P_2$] (q2) at (0.9,0) {};
  \node[htrans,label=right:{\small$T_1$}] (u) at (0,-1.3) {};
  \node[place,label=below:$P_3$] (q3) at (-1.3,-2.7) {\tok\tok\tok};
  \node[place,label=below:$P_4$] (q4) at (0,-2.7) {\tok};
  \node[place,label=below:$P_5$] (q5) at (1.3,-2.7) {\tok\tok};
  \draw[arc] (q1) -- (u);
  \draw[arc] (q2) -- node[wt,right]{2} (u);
  \draw[arc] (u) -- node[wt,left]{3} (q3);
  \draw[arc] (u) -- (q4);
  \draw[arc] (u) -- node[wt,right]{2} (q5);
  \end{scope}
\end{tikzpicture}
\caption{Firing of a transition. Unlabelled arcs have weight 1. Firing $T_1$ changes the marking $P_1+2P_2$ to $3P_3+P_4+2P_5$.}\label{fig:firing}
\end{figure}

\subsection{Elementary regulatory modules}
Petri nets have been used to model GRNs in several ways \citep{Chaouiya2004,Chaouiya2006,Chaouiya2007,Chaouiya2008,LiuHeiner2014,LiuHeinerGilbert2019,LiuHeinerGilbert2020,LiuHC2017,LiuChen2018}. Here we adopt a simple encoding based on \citet{Chaouiya2004}, \citet{Bosnacki2004} and \citet{Ruths2008}. Genes (more precisely, their products) are places, and tokens in a place represent units of the corresponding gene product. In a GRN diagram, an arrow from gene $g_2$ to gene $g_1$ means that $g_2$ activates $g_1$ (Figure~\ref{fig:activation}a), and a blunt-ended line means that $g_2$ inhibits $g_1$ (Figure~\ref{fig:inhibition}a). Each regulatory interaction becomes a transition $\T{g_1}{g_2}$ (target first, regulator second).

In each case the regulator $g_2$ may or may not be consumed. A double-headed arrow between $g_2$ and the transition means that $g_2$ is both an input and an output of the transition, so firing leaves the number of $g_2$ tokens unchanged (a \emph{read arc}, or self-loop); a single arrow means that $g_2$ is consumed. In terms of Definition~\ref{def:petri}, the four modules of Figures~\ref{fig:activation} and~\ref{fig:inhibition} are:

\begin{center}
\renewcommand{\arraystretch}{1.25}
\begin{tabular}{@{}llll@{}}
\toprule
Interaction & Figure & $s(\T{g_1}{g_2})$ & $t(\T{g_1}{g_2})$ \\
\midrule
activation, $g_2$ consumed & \ref{fig:activation}b & $g_2$ & $g_1$ \\
activation, $g_2$ not consumed & \ref{fig:activation}c & $g_2$ & $g_1+g_2$ \\
inhibition, $g_2$ consumed & \ref{fig:inhibition}b & $g_1+g_2$ & $0$ \\
inhibition, $g_2$ not consumed & \ref{fig:inhibition}c & $g_1+g_2$ & $g_2$ \\
\midrule
degradation of $g$ & --- & $g$ & $0$ \\
\bottomrule
\end{tabular}
\end{center}

\noindent Thus activation produces a token of $g_1$, while inhibition removes one: in this encoding, inhibition of $g_1$ by $g_2$ is modelled as $g_2$-dependent removal of the product of $g_1$ (for example by sequestration or targeted degradation). The last row adds a \emph{degradation} transition $d_g$ for a gene $g$, which removes one token of $g$; we return to its role in Remark~\ref{rem:limitations}.

\begin{figure}[t]
\centering
\begin{tikzpicture}[scale=0.9, every node/.append style={transform shape}]
  \node[font=\small] at (-0.6,0.6) {(a)};
  \node[pt,label=above:{\small$g_2$}] (a2) at (0.3,0) {};
  \node[pt,label=above:{\small$g_1$}] (a1) at (1.5,0) {};
  \draw[arc] (a2) -- (a1);
  \node[font=\small] at (-0.6,-1.0) {(b)};
  \begin{scope}[yshift=-1.9cm]
    \node[place,label=above:$g_2$] (b2) at (0.3,0) {\tok};
    \node[vtrans,label=above:{\scriptsize$\T{g_1}{g_2}$}] (bt) at (1.5,0) {};
    \node[place,label=above:$g_1$] (b1) at (2.7,0) {};
    \draw[arc] (b2) -- (bt); \draw[arc] (bt) -- (b1);
    \draw[arc] (3.6,0) -- node[above,font=\scriptsize]{$\T{g_1}{g_2}$ fires} (5.4,0);
    \node[place,label=above:$g_2$] (c2) at (6.3,0) {};
    \node[vtrans,label=above:{\scriptsize$\T{g_1}{g_2}$}] (ct) at (7.5,0) {};
    \node[place,label=above:$g_1$] (c1) at (8.7,0) {\tok};
    \draw[arc] (c2) -- (ct); \draw[arc] (ct) -- (c1);
  \end{scope}
  \node[font=\small] at (-0.6,-3.2) {(c)};
  \begin{scope}[yshift=-4.1cm]
    \node[place,label=above:$g_2$] (b2) at (0.3,0) {\tok};
    \node[vtrans,label=above:{\scriptsize$\T{g_1}{g_2}$}] (bt) at (1.5,0) {};
    \node[place,label=above:$g_1$] (b1) at (2.7,0) {};
    \draw[rarc] (b2) -- (bt); \draw[arc] (bt) -- (b1);
    \draw[arc] (3.6,0) -- node[above,font=\scriptsize]{$\T{g_1}{g_2}$ fires} (5.4,0);
    \node[place,label=above:$g_2$] (c2) at (6.3,0) {\tok};
    \node[vtrans,label=above:{\scriptsize$\T{g_1}{g_2}$}] (ct) at (7.5,0) {};
    \node[place,label=above:$g_1$] (c1) at (8.7,0) {\tok};
    \draw[rarc] (c2) -- (ct); \draw[arc] (ct) -- (c1);
  \end{scope}
\end{tikzpicture}
\caption{Activation of gene $g_1$ by $g_2$. (a) GRN notation. (b) Activation in which $g_2$ is consumed. (c) Activation in which $g_2$ is not consumed (read arc).}\label{fig:activation}
\end{figure}

\begin{figure}[t]
\centering
\begin{tikzpicture}[scale=0.9, every node/.append style={transform shape}]
  \node[font=\small] at (-0.6,0.6) {(a)};
  \node[pt,label=above:{\small$g_2$}] (a2) at (0.3,0) {};
  \node[pt,label=above:{\small$g_1$}] (a1) at (1.5,0) {};
  \draw[-{Bar[width=2.5mm]},thick] (a2) -- (a1);
  \node[font=\small] at (-0.6,-1.0) {(b)};
  \begin{scope}[yshift=-1.9cm]
    \node[place,label=above:$g_2$] (b2) at (0.3,0) {\tok};
    \node[vtrans,label=above:{\scriptsize$\T{g_1}{g_2}$}] (bt) at (1.5,0) {};
    \node[place,label=above:$g_1$] (b1) at (2.7,0) {\tok};
    \draw[arc] (b2) -- (bt); \draw[arc] (b1) -- (bt);
    \draw[arc] (3.6,0) -- node[above,font=\scriptsize]{$\T{g_1}{g_2}$ fires} (5.4,0);
    \node[place,label=above:$g_2$] (c2) at (6.3,0) {};
    \node[vtrans,label=above:{\scriptsize$\T{g_1}{g_2}$}] (ct) at (7.5,0) {};
    \node[place,label=above:$g_1$] (c1) at (8.7,0) {};
    \draw[arc] (c2) -- (ct); \draw[arc] (c1) -- (ct);
  \end{scope}
  \node[font=\small] at (-0.6,-3.2) {(c)};
  \begin{scope}[yshift=-4.1cm]
    \node[place,label=above:$g_2$] (b2) at (0.3,0) {\tok};
    \node[vtrans,label=above:{\scriptsize$\T{g_1}{g_2}$}] (bt) at (1.5,0) {};
    \node[place,label=above:$g_1$] (b1) at (2.7,0) {\tok};
    \draw[rarc] (b2) -- (bt); \draw[arc] (b1) -- (bt);
    \draw[arc] (3.6,0) -- node[above,font=\scriptsize]{$\T{g_1}{g_2}$ fires} (5.4,0);
    \node[place,label=above:$g_2$] (c2) at (6.3,0) {\tok};
    \node[vtrans,label=above:{\scriptsize$\T{g_1}{g_2}$}] (ct) at (7.5,0) {};
    \node[place,label=above:$g_1$] (c1) at (8.7,0) {};
    \draw[rarc] (c2) -- (ct); \draw[arc] (c1) -- (ct);
  \end{scope}
\end{tikzpicture}
\caption{Inhibition of gene $g_1$ by $g_2$. (a) GRN notation. (b) Inhibition in which $g_2$ is consumed. (c) Inhibition in which $g_2$ is not consumed.}\label{fig:inhibition}
\end{figure}

\subsection{From a GRN to a Petri net}
With these building blocks, a GRN can be translated into a Petri net interaction by interaction. Consider the GRN in Figure~\ref{fig:grn}a: gene $A$ activates genes $B$ and $C$, gene $B$ activates $D$, and gene $C$ inhibits $D$ and activates $E$. Suppose that $C$ is consumed when it activates $E$ and that no other regulator is consumed. The corresponding Petri net (Figure~\ref{fig:grn}b) has places $\{A,B,C,D,E\}$ and transitions
\begin{equation}\label{eq:grn-transitions}
\begin{aligned}
\T{B}{A}&\colon A\to A+B, & \T{C}{A}&\colon A\to A+C, & \T{D}{B}&\colon B\to B+D,\\
\T{D}{C}&\colon C+D\to C, & \T{E}{C}&\colon C\to E,
\end{aligned}
\end{equation}
where $\tau\colon a\to b$ means $s(\tau)=a$ and $t(\tau)=b$. Once a GRN is written as a Petri net, standard Petri net tools can be used to simulate and analyse it \citep{Ruths2008,LiuHeiner2014}.

\begin{figure}[t]
\centering
\begin{tikzpicture}
  \node[font=\small] at (-1.0,1.2) {(a)};
  \node[lab] (A) at (0.8,1.0) {$A$};
  \node[lab] (B) at (0.1,0.2) {$B$};
  \node[lab] (C) at (1.5,0.2) {$C$};
  \node[lab] (D) at (0.8,-0.6) {$D$};
  \node[lab] (E) at (2.2,-0.6) {$E$};
  \draw[arc] (A) -- (B); \draw[arc] (A) -- (C);
  \draw[arc] (B) -- (D); \draw[-{Bar[width=2.5mm]},thick] (C) -- (D);
  \draw[arc] (C) -- (E);
  \begin{scope}[xshift=4.2cm, yshift=1.0cm]
  \node[font=\small] at (-1.0,0.2) {(b)};
  \node[place] (pA) at (0,0) {$A$};
  \node[vtrans,label=above:{\scriptsize$\T{C}{A}$}] (tCA) at (1.4,0) {};
  \node[place] (pC) at (2.8,0) {$C$};
  \node[vtrans,label=above:{\scriptsize$\T{E}{C}$}] (tEC) at (4.2,0) {};
  \node[place] (pE) at (5.6,0) {$E$};
  \node[htrans,label=left:{\scriptsize$\T{B}{A}$}] (tBA) at (0,-1.3) {};
  \node[htrans,label=right:{\scriptsize$\T{D}{C}$}] (tDC) at (2.8,-1.3) {};
  \node[place] (pB) at (0,-2.6) {$B$};
  \node[vtrans,label=below:{\scriptsize$\T{D}{B}$}] (tDB) at (1.4,-2.6) {};
  \node[place] (pD) at (2.8,-2.6) {$D$};
  \draw[rarc] (pA) -- (tCA); \draw[arc] (tCA) -- (pC);
  \draw[arc] (pC) -- (tEC); \draw[arc] (tEC) -- (pE);
  \draw[rarc] (pA) -- (tBA); \draw[arc] (tBA) -- (pB);
  \draw[rarc] (pB) -- (tDB); \draw[arc] (tDB) -- (pD);
  \draw[rarc] (pC) -- (tDC); \draw[arc] (pD) -- (tDC);
  \end{scope}
\end{tikzpicture}
\caption{(a) A simple GRN. (b) Its Petri net representation, with transitions given in \eqref{eq:grn-transitions}. Only $C$ is consumed (when it activates $E$).}\label{fig:grn}
\end{figure}

\begin{remark}[Limitations of the encoding]\label{rem:limitations}
Three points are worth keeping in mind.
\begin{enumerate}[nosep,label=(\roman*)]
\item \emph{Unbounded accumulation.} A non-consuming activation such as $\T{B}{A}\colon A\to A+B$ remains enabled for as long as $A$ is present, so it can fire indefinitely and $B$ accumulates without bound. Real gene products decay, and adding a degradation transition $d_g\colon g\to 0$ for each gene restores this. Its effect is most visible in the rate-equation semantics of Section~\ref{sec:grn-dynamics}, where it produces the familiar linear decay terms of ODE models of GRNs.
\item \emph{Inhibitor arcs.} Some Petri net models of regulation use inhibitor arcs, which allow a transition to fire only when a place is \emph{empty}. A transition with an inhibitor arc cannot be described by source and target functions $s,t\colon T\to\Nset{S}$, so it lies outside Definition~\ref{def:petri} and outside the open Petri net framework of Section~\ref{sec:open}. The modules above avoid inhibitor arcs, as do translations of logical regulatory graphs that use complementary places instead \citep{Chaouiya2004}.
\item \emph{Granularity.} Tokens here are abstract units of gene product, and the encoding does not distinguish transcription from translation or model promoter states. Finer models can be written in the same language by adding places (for example, for promoter states or mRNA and protein separately) and transitions between them.
\end{enumerate}
\end{remark}

\section{Petri nets as monoids: an algebra of processes}\label{sec:monoids}
\citet{MeseguerMontanari1990} observed that a Petri net is a directed graph whose set of nodes carries the structure of a free commutative monoid, and that the behaviour of the net is then described by composing transitions in two ways, sequentially and in parallel. We recall this point of view, because it underlies the operational semantics of open Petri nets (Section~\ref{sec:semantics}), and because it is important to be clear about what it composes: \emph{processes inside one net}, not separate nets.

\subsection{Two equivalent definitions}
A traditional definition of a Petri net is a triple $(S,T,F)$ where $S$ is a set of places, $T$ a set of transitions, and $F\colon(S\times T)+(T\times S)\to\N$ is a function giving the weight of each arc ($+$ denotes disjoint union and $F=0$ means no arc). Equivalently, $F$ is a multiset of arcs. Meseguer and Montanari instead start from graphs:

\begin{definition}
A \emph{graph} consists of a set $T$ of arcs, a set $V$ of nodes and two functions $\partial_0,\partial_1\colon T\to V$, called source and target. A \emph{graph morphism} from $(T,V,\partial_0,\partial_1)$ to $(T',V',\partial'_0,\partial'_1)$ is a pair of functions $f\colon T\to T'$, $g\colon V\to V'$ such that $g\circ\partial_0=\partial'_0\circ f$ and $g\circ\partial_1=\partial'_1\circ f$.
\end{definition}

A Petri net is then a graph whose arcs are the transitions and whose set of nodes is the free commutative monoid $\Nset{S}$ on the set $S$ of places (written $S^{\oplus}$ by Meseguer and Montanari). This is exactly Definition~\ref{def:petri} with $\partial_0=s$ and $\partial_1=t$. The two descriptions are related by
\[
s(\tau)=\sum_{p\in S}F(p,\tau)\,p,\qquad t(\tau)=\sum_{p\in S}F(\tau,p)\,p .
\]
For example, the net in Figure~\ref{fig:smallnet}, with $S=\{g_1,g_2,g_3\}$, $T=\{t\}$ and $F(g_1,t)=F(g_2,t)=F(t,g_1)=F(t,g_3)=1$, has $s(t)=g_1+g_2$ and $t(t)=g_1+g_3$. (In GRN terms: $g_1$ and $g_2$ jointly produce $g_3$; $g_1$ is not consumed but $g_2$ is.) We write the monoid operation of $\Nset{S}$ as $+$ throughout.

\paragraph{Morphisms.} Two different notions of morphism appear in the literature, and the difference matters for composition. \citet{MeseguerMontanari1990} allow a morphism of Petri nets to act on markings by an arbitrary monoid homomorphism $\Nset{S}\to\Nset{S'}$, so that one place may be sent to a formal sum of places. \citet{BaezMaster2020}, following \citet{Baldan2005} and \citet{Baldan2015}, use a narrower notion:

\begin{definition}[\citealp{BaezMaster2020}]\label{def:petri-morphism}
A \emph{morphism} from $(s,t\colon T\to\Nset{S})$ to $(s',t'\colon T'\to\Nset{S'})$ is a pair of functions $f\colon T\to T'$ and $g\colon S\to S'$ such that $s'\circ f=\Nset{g}\circ s$ and $t'\circ f=\Nset{g}\circ t$, where $\Nset{g}\colon\Nset{S}\to\Nset{S'}$ is the monoid homomorphism extending $g$. Petri nets and these morphisms form a category $\Petri$.
\end{definition}

So morphisms in $\Petri$ send places to places. This is what makes the gluing of nets along places in Section~\ref{sec:open} well behaved.

\begin{figure}[t]
\centering
\begin{tikzpicture}
  \node[place,label=above:$g_1$] (g1) at (-0.9,0) {};
  \node[place,label=above:$g_2$] (g2) at (0.9,0) {};
  \node[htrans,label=right:{\small$t$}] (t) at (0,-1.3) {};
  \node[place,label=below:$g_3$] (g3) at (0,-2.6) {};
  \draw[rarc] (g1) -- (t); \draw[arc] (g2) -- (t); \draw[arc] (t) -- (g3);
\end{tikzpicture}
\caption{A Petri net with $s(t)=g_1+g_2$ and $t(t)=g_1+g_3$.}\label{fig:smallnet}
\end{figure}

\subsection{Processes and their composition}
Given a Petri net $P$, \citet{MeseguerMontanari1990} construct a monoidal category of its processes; we follow the presentation of \citet[Sec.~2]{BaezMaster2020}, who call it $FP$. Its objects are markings $m\in\Nset{S}$. Its morphisms are generated by
\begin{itemize}[nosep]
\item a morphism $\tau\colon s(\tau)\to t(\tau)$ for each transition $\tau$,
\item an identity $\id_m\colon m\to m$ for each marking $m$,
\item \emph{parallel composition}: for $\alpha\colon a\to b$ and $\beta\colon a'\to b'$, a morphism $\alpha\oplus\beta\colon a+a'\to b+b'$,
\item \emph{sequential composition}: for $\alpha\colon a\to b$ and $\beta\colon b\to c$, a morphism $\beta\circ\alpha\colon a\to c$,
\end{itemize}
subject to the laws of a commutative monoidal category, i.e.\ a strict monoidal category in which $\oplus$ is commutative on both objects and morphisms. A morphism $m\to n$ describes a way of getting from marking $m$ to marking $n$ by firing transitions, some possibly in parallel. In particular, $n$ is reachable from $m$ by a finite sequence of firings if and only if there is a morphism $m\to n$ in $FP$ \citep[Prop.~21]{BaezMaster2020}. The assignment $P\mapsto FP$ extends to a functor $F\colon\Petri\to\CMC$ into the category of commutative monoidal categories, and this functor is a left adjoint \citep[Lemma~9]{BaezMaster2020}.

\begin{example}[Sequential composition]\label{ex:sequential}
Let $N$ be the net of Figure~\ref{fig:sequential}a, with transitions $t\colon g_1+g_2\to g_1+g_3$ and $t'\colon g_3\to g_4$. The two transitions cannot be composed directly, because the target $g_1+g_3$ of $t$ is not the source $g_3$ of $t'$. Composing $t'$ in parallel with the identity on $g_1$ fixes this, giving
\[
(\id_{g_1}\oplus t')\circ t\colon\; g_1+g_2\;\longrightarrow\; g_1+g_4 ,
\]
the process ``fire $t$, then fire $t'$''. Figure~\ref{fig:sequential}b draws this process as a single derived transition (dashed). It is a morphism of $FN$, not a transition of a new Petri net: no new net is built.
\end{example}

\begin{figure}[t]
\centering
\begin{tikzpicture}
  \node[font=\small] at (-1.8,0.5) {(a)};
  \node[place,label=above:$g_1$] (g1) at (-0.9,0) {};
  \node[place,label=above:$g_2$] (g2) at (0.9,0) {};
  \node[htrans,label=right:{\small$t$}] (t) at (0,-1.2) {};
  \node[place,label=right:$g_3$] (g3) at (0,-2.4) {};
  \node[htrans,label=right:{\small$t'$}] (tp) at (0,-3.6) {};
  \node[place,label=below:$g_4$] (g4) at (0,-4.8) {};
  \draw[rarc] (g1) -- (t); \draw[arc] (g2) -- (t); \draw[arc] (t) -- (g3);
  \draw[arc] (g3) -- (tp); \draw[arc] (tp) -- (g4);
  \begin{scope}[xshift=5.5cm]
  \node[font=\small] at (-2.0,0.5) {(b)};
  \node[place,label=above:$g_1$] (h1) at (-0.9,0) {};
  \node[place,label=above:$g_2$] (h2) at (0.9,0) {};
  \node[dtrans,label=right:{\small$(\id_{g_1}\oplus t')\circ t$}] (d) at (0,-2.4) {};
  \node[place,label=below:$g_4$] (h4) at (0,-4.8) {};
  \draw[rarc] (h1) -- (d); \draw[arc] (h2) -- (d); \draw[arc] (d) -- (h4);
  \end{scope}
\end{tikzpicture}
\caption{Sequential composition of processes. (a) A net $N$ with transitions $t$ and $t'$. (b) The composite process $(\id_{g_1}\oplus t')\circ t\colon g_1+g_2\to g_1+g_4$ of $N$, drawn as a derived (dashed) transition.}\label{fig:sequential}
\end{figure}

\begin{example}[Parallel composition]\label{ex:parallel}
Let $N_1$ have one transition $t\colon 2g_1\to g_2$ and let $N_2$ have one transition $t'\colon g'_1+2g'_2\to g'_3$ (Figure~\ref{fig:parallel}a,b). Running both nets side by side gives their disjoint union $N_1+N_2$, which is the coproduct in $\Petri$; its places are $\{g_1,g_2,g'_1,g'_2,g'_3\}$ and its transitions are $t$ and $t'$. In $F(N_1+N_2)$ the parallel composite
\[
t\oplus t'\colon\; 2g_1+g'_1+2g'_2\;\longrightarrow\; g_2+g'_3
\]
describes firing $t$ and $t'$ concurrently (Figure~\ref{fig:parallel}c). The places are those of $N_1$ and $N_2$ unchanged, and the arc weights are those of $t$ and $t'$ unchanged: no new places such as ``$g_1\otimes g'_1$'' are formed, and weights are not multiplied.
\end{example}

\begin{figure}[t]
\centering
\begin{tikzpicture}
  \node[font=\small] at (-0.9,0.6) {(a)};
  \node[place,label=above:$g_1$] (a1) at (0,0) {};
  \node[htrans,label=right:{\small$t$}] (at) at (0,-1.3) {};
  \node[place,label=below:$g_2$] (a2) at (0,-2.6) {};
  \draw[arc] (a1) -- node[wt,right]{2} (at); \draw[arc] (at) -- (a2);
  \begin{scope}[xshift=2.6cm]
  \node[font=\small] at (-1.3,0.6) {(b)};
  \node[place,label=above:$g'_1$] (b1) at (-0.6,0) {};
  \node[place,label=above:$g'_2$] (b2) at (0.6,0) {};
  \node[htrans,label=right:{\small$t'$}] (bt) at (0,-1.3) {};
  \node[place,label=below:$g'_3$] (b3) at (0,-2.6) {};
  \draw[arc] (b1) -- (bt); \draw[arc] (b2) -- node[wt,right]{2} (bt); \draw[arc] (bt) -- (b3);
  \end{scope}
  \begin{scope}[xshift=7.2cm]
  \node[font=\small] at (-2.5,0.6) {(c)};
  \node[place,label=above:$g_1$] (c1) at (-1.6,0) {};
  \node[place,label=above:$g'_1$] (c2) at (0,0) {};
  \node[place,label=above:$g'_2$] (c3) at (1.6,0) {};
  \node[dtrans,label=right:{\small$t\oplus t'$}] (ct) at (0,-1.3) {};
  \node[place,label=below:$g_2$] (c4) at (-0.8,-2.6) {};
  \node[place,label=below:$g'_3$] (c5) at (0.8,-2.6) {};
  \draw[arc] (c1) -- node[wt,above left]{2} (ct); \draw[arc] (c2) -- (ct);
  \draw[arc] (c3) -- node[wt,above right]{2} (ct);
  \draw[arc] (ct) -- (c4); \draw[arc] (ct) -- (c5);
  \end{scope}
\end{tikzpicture}
\caption{Parallel composition. (a) $N_1$. (b) $N_2$. (c) The process $t\oplus t'\colon 2g_1+g'_1+2g'_2\to g_2+g'_3$ of the disjoint union $N_1+N_2$, drawn as a derived (dashed) transition.}\label{fig:parallel}
\end{figure}

\subsection{What the monoidal view does not provide}
Sequential and parallel composition of processes describe the \emph{behaviour} of a given net. The only way of combining two nets in this framework is the disjoint union $N_1+N_2$, in which the two nets do not interact. A GRN, however, is assembled from sub-networks that \emph{share genes}: the target of one regulatory module is a regulator in another. What is needed is a way to glue nets together along common places, and to do so in a way that is compatible with the process semantics above. This is what open Petri nets provide.

\section{Open Petri nets}\label{sec:open}
We follow \citet{BaezMaster2020} throughout this section.

\subsection{Definition}
Let $L\colon\Set\to\Petri$ send a set $X$ to the Petri net with set of places $X$ and no transitions (and a function to the evident morphism). $L$ is left adjoint to the functor $R\colon\Petri\to\Set$ that sends a Petri net to its set of places \citep[Lemma~11]{BaezMaster2020}. Consequently, a morphism of Petri nets $LX\to P$ is the same as a function from $X$ to the set of places of $P$.

\begin{definition}[\citealp{BaezMaster2020}]
An \emph{open Petri net} from a set $X$ to a set $Y$, written $P\colon X\pto Y$, is a cospan in $\Petri$ of the form
\[
\begin{tikzcd}[column sep=small]
 & P & \\
LX \arrow[ur,"i"] & & LY \arrow[ul,"o"']
\end{tikzcd}
\]
or equivalently a Petri net $P$ with set of places $S$ together with functions $i\colon X\to S$ and $o\colon Y\to S$.
\end{definition}

The functions $i$ and $o$ mark certain places as \emph{inputs} and \emph{outputs}: points at which tokens may flow in or out, making the net ``open''. They need not be injective, and a place may be both an input and an output. The words ``input'' and ``output'' only fix the direction in which open nets are composed; they do not restrict the direction in which tokens move across a boundary.

\subsection{Composition by pushout}\label{sec:pushout}
Given open Petri nets $P\colon X\pto Y$ and $Q\colon Y\pto Z$, their composite $Q\odot P\colon X\pto Z$ is obtained by gluing the outputs of $P$ to the inputs of $Q$. Formally, it is the composite of cospans computed by a pushout in $\Petri$ \citep[Thm.~13]{BaezMaster2020}:
\[
\begin{tikzcd}[column sep=small,row sep=small]
 & & P+_{LY}Q & & \\
 & P \arrow[ur,"j_P"] & & Q \arrow[ul,"j_Q"'] & \\
LX \arrow[ur,"i_P"] & & LY \arrow[ul,"o_P"'] \arrow[ur,"i_Q"] & & LZ \arrow[ul,"o_Q"']
\end{tikzcd}
\]
with legs $j_P\circ i_P\colon LX\to P+_{LY}Q$ and $j_Q\circ o_Q\colon LZ\to P+_{LY}Q$. Because $LY$ has no transitions and colimits in $\Petri$ are computed on places and on transitions separately \citep[Lemma~15]{BaezMaster2020}, the pushout can be described explicitly. Let $P$ have places $S_P$ and transitions $T_P$, and $Q$ have places $S_Q$ and transitions $T_Q$. Then:
\begin{itemize}[nosep]
\item the places of $P+_{LY}Q$ are $S=(S_P+S_Q)/{\sim}$, where $\sim$ is the smallest equivalence relation with $o_P(y)\sim i_Q(y)$ for all $y\in Y$; let $q\colon S_P+S_Q\to S$ be the quotient map;
\item the transitions are $T_P+T_Q$, with source and target obtained by applying $\Nset{q}$, e.g.\ $s(\tau)=\Nset{q}(s_P(\tau))$ for $\tau\in T_P$;
\item the input and output maps are $q\circ i_P\colon X\to S$ and $q\circ o_Q\colon Z\to S$.
\end{itemize}
Three consequences of this description are important in practice:
\begin{enumerate}[nosep,label=(\alph*)]
\item \emph{Arc weights add.} If a transition has arcs to two places that become identified, the composite has a single arc whose weight is the sum, since $\Nset{q}(p+p')=2\,q(p)$ when $q(p)=q(p')$.
\item \emph{Transitions are never identified.} The transitions of $Q\odot P$ are exactly those of $P$ together with those of $Q$. A transition that appears in both $P$ and $Q$ therefore appears twice in the composite.
\item \emph{Composition is defined only up to isomorphism.} A merged place is an equivalence class of places; giving it a name, e.g.\ one of the original names, is a choice, and different choices give isomorphic nets.
\end{enumerate}

\begin{example}[A toy composition]\label{ex:toy}
The following example is adapted from an early version of \citet{BaezMaster2020}; its places and transitions have no particular biological meaning. Let $P\colon X\pto Y$ have places $p_1,\dots,p_5$ and one transition $t\colon p_1+p_2\to p_1+p_3+p_4+p_5$, with $X=\{1\}$, $i(1)=p_1$, $Y=\{2,3,4\}$ and $o(2)=p_3$, $o(3)=p_4$, $o(4)=p_5$. Let $Q\colon Y\pto Z$ have places $p_6,p_7,p_8$ and one transition $t'\colon p_6+p_7+p_8\to p_6$, with $i(2)=i(3)=p_6$, $i(4)=p_7$, $Z=\{5\}$ and $o(5)=p_8$ (Figure~\ref{fig:toy}a). To compose them we first place them side by side, sharing $Y$ (Figure~\ref{fig:toy}b). Next we identify places that are images of the same point of $Y$: the point $2$ gives $p_3\sim p_6$, and the point $3$ gives $p_4\sim p_6$, so $p_3$, $p_4$ and $p_6$ form one class; the point $4$ gives $p_5\sim p_7$. Naming the merged places $p_6$ and $p_7$ and erasing $Y$, we obtain $Q\odot P\colon X\pto Z$ (Figure~\ref{fig:toy}c) with
\[
t\colon p_1+p_2\to p_1+2p_6+p_7,\qquad t'\colon p_6+p_7+p_8\to p_6 .
\]
Note the arc of weight $2$ from $t$ to $p_6$, which arises because both $p_3$ and $p_4$ were merged into $p_6$ (consequence (a) above).
\end{example}

\begin{figure}[htbp]
\centering
\begin{tikzpicture}[scale=0.85, every node/.append style={transform shape}]
  \node[font=\small] at (-1.2,1.6) {(a)};
  \node[bnd,minimum height=3.4cm,label=above:$X$] (X) at (0,0) {};
  \node[pt,label={[font=\small,label distance=2.4mm]left:1}] (x1) at (0,0) {};
  \node[place] (p1) at (1.4,0) {$p_1$};
  \node[place] (p2) at (1.4,-1.2) {$p_2$};
  \node[htrans] (t) at (2.9,0) {$t$};
  \node[place] (p3) at (4.4,1.1) {$p_3$};
  \node[place] (p4) at (4.4,0) {$p_4$};
  \node[place] (p5) at (4.4,-1.1) {$p_5$};
  \node[bnd,minimum height=3.4cm,label=above:$Y$] (Y) at (5.8,0) {};
  \node[pt,label={[font=\small,label distance=2.4mm]right:2}] (y2) at (5.8,1.1) {};
  \node[pt,label={[font=\small,label distance=2.4mm]right:3}] (y3) at (5.8,0) {};
  \node[pt,label={[font=\small,label distance=2.4mm]right:4}] (y4) at (5.8,-1.1) {};
  \draw[leg] (x1) -- (p1);
  \draw[rarc] (p1) -- (t); \draw[arc] (p2) -- (t);
  \draw[arc] (t) -- (p3); \draw[arc] (t) -- (p4); \draw[arc] (t) -- (p5);
  \draw[leg] (y2) -- (p3); \draw[leg] (y3) -- (p4); \draw[leg] (y4) -- (p5);
  \begin{scope}[xshift=8.0cm]
  \node[bnd,minimum height=3.4cm,label=above:$Y$] (Yb) at (0,0) {};
  \node[pt,label={[font=\small,label distance=2.4mm]left:2}] (z2) at (0,1.1) {};
  \node[pt,label={[font=\small,label distance=2.4mm]left:3}] (z3) at (0,0) {};
  \node[pt,label={[font=\small,label distance=2.4mm]left:4}] (z4) at (0,-1.1) {};
  \node[place] (p6) at (1.5,0.4) {$p_6$};
  \node[place] (p7) at (1.5,-1.1) {$p_7$};
  \node[htrans] (tp) at (3.0,0.4) {$t'$};
  \node[place] (p8) at (4.4,0.4) {$p_8$};
  \node[bnd,minimum height=3.4cm,label=above:$Z$] (Z) at (5.6,0) {};
  \node[pt,label={[font=\small,label distance=2.4mm]right:5}] (z5) at (5.6,0.4) {};
  \draw[leg] (z2) -- (p6); \draw[leg] (z3) -- (p6); \draw[leg] (z4) -- (p7);
  \draw[rarc] (p6) -- (tp); \draw[arc] (p7) -- (tp); \draw[arc] (p8) -- (tp);
  \draw[leg] (z5) -- (p8);
  \end{scope}
  \begin{scope}[yshift=-4.6cm, xshift=1.0cm]
  \node[font=\small] at (-2.2,1.6) {(b)};
  \node[bnd,minimum height=3.4cm,label=above:$X$] (X) at (0,0) {};
  \node[pt,label={[font=\small,label distance=2.4mm]left:1}] (x1) at (0,0) {};
  \node[place] (p1) at (1.4,0) {$p_1$};
  \node[place] (p2) at (1.4,-1.2) {$p_2$};
  \node[htrans] (t) at (2.9,0) {$t$};
  \node[place] (p3) at (4.4,1.1) {$p_3$};
  \node[place] (p4) at (4.4,0) {$p_4$};
  \node[place] (p5) at (4.4,-1.1) {$p_5$};
  \node[bnd,minimum height=3.4cm,label=above:$Y$] (Y) at (5.8,0) {};
  \node[pt] (y2) at (5.8,1.1) {}; \node[font=\scriptsize] at (5.95,1.35) {2};
  \node[pt] (y3) at (5.8,0) {}; \node[font=\scriptsize] at (5.95,0.25) {3};
  \node[pt] (y4) at (5.8,-1.1) {}; \node[font=\scriptsize] at (5.95,-0.85) {4};
  \node[place] (p6) at (7.3,0.4) {$p_6$};
  \node[place] (p7) at (7.3,-1.1) {$p_7$};
  \node[htrans] (tp) at (8.8,0.4) {$t'$};
  \node[place] (p8) at (10.2,0.4) {$p_8$};
  \node[bnd,minimum height=3.4cm,label=above:$Z$] (Z) at (11.4,0) {};
  \node[pt,label={[font=\small,label distance=2.4mm]right:5}] (z5) at (11.4,0.4) {};
  \draw[leg] (x1) -- (p1);
  \draw[rarc] (p1) -- (t); \draw[arc] (p2) -- (t);
  \draw[arc] (t) -- (p3); \draw[arc] (t) -- (p4); \draw[arc] (t) -- (p5);
  \draw[leg] (y2) -- (p3); \draw[leg] (y3) -- (p4); \draw[leg] (y4) -- (p5);
  \draw[leg] (y2) -- (p6); \draw[leg] (y3) -- (p6); \draw[leg] (y4) -- (p7);
  \draw[rarc] (p6) -- (tp); \draw[arc] (p7) -- (tp); \draw[arc] (p8) -- (tp);
  \draw[leg] (z5) -- (p8);
  \end{scope}
  \begin{scope}[yshift=-9.2cm, xshift=2.4cm]
  \node[font=\small] at (-2.2,1.6) {(c)};
  \node[bnd,minimum height=3.4cm,label=above:$X$] (X) at (0,0) {};
  \node[pt,label={[font=\small,label distance=2.4mm]left:1}] (x1) at (0,0) {};
  \node[place] (p1) at (1.4,0) {$p_1$};
  \node[place] (p2) at (1.4,-1.2) {$p_2$};
  \node[htrans] (t) at (2.9,0) {$t$};
  \node[place] (p6) at (4.6,0.9) {$p_6$};
  \node[place] (p7) at (4.6,-0.9) {$p_7$};
  \node[htrans] (tp) at (6.2,0.0) {$t'$};
  \node[place] (p8) at (7.6,0.0) {$p_8$};
  \node[bnd,minimum height=3.4cm,label=above:$Z$] (Z) at (8.8,0) {};
  \node[pt,label={[font=\small,label distance=2.4mm]right:5}] (z5) at (8.8,0) {};
  \draw[leg] (x1) -- (p1);
  \draw[rarc] (p1) -- (t); \draw[arc] (p2) -- (t);
  \draw[arc] (t) -- node[wt,above left]{2} (p6); \draw[arc] (t) -- (p7);
  \draw[rarc] (p6) -- (tp); \draw[arc] (p7) -- (tp); \draw[arc] (p8) -- (tp);
  \draw[leg] (z5) -- (p8);
  \end{scope}
\end{tikzpicture}
\caption{Composition of open Petri nets (Example~\ref{ex:toy}). Dashed grey arrows are the input and output maps. (a) $P\colon X\pto Y$ and $Q\colon Y\pto Z$. (b) The two nets placed side by side along $Y$. (c) The composite $Q\odot P\colon X\pto Z$; the arc from $t$ to $p_6$ has weight $2$.}\label{fig:toy}
\end{figure}

\subsection{The double category $\DOpenPetri$}
Since the pushout is defined only up to isomorphism, $\odot$ is associative only up to isomorphism. One way to obtain a category is to use \emph{isomorphism classes} of open Petri nets as morphisms; this gives a symmetric monoidal category $\OpenPetri$ whose objects are sets and whose tensor product is disjoint union \citep{BaezMaster2020}. Passing to isomorphism classes, however, means that one can no longer point to a particular place or transition of a composite. For GRNs this is a real loss: we want to be able to say which place of the composite network is gene $B$. \citet{BaezMaster2020} therefore work with a symmetric monoidal \emph{double category} $\DOpenPetri$, constructed using the structured cospan machinery of \citet{BaezCourser2020}. It has:
\begin{enumerate}[nosep,label=(\arabic*)]
\item \emph{objects}: sets $X,Y,Z,\dots$;
\item \emph{vertical 1-morphisms}: functions $f\colon X\to X'$;
\item \emph{horizontal 1-cells}: open Petri nets $P\colon X\pto Y$;
\item \emph{2-morphisms}: given open Petri nets $LX\xrightarrow{i}P\xleftarrow{o}LY$ and $LX'\xrightarrow{i'}P'\xleftarrow{o'}LY'$ and functions $f\colon X\to X'$, $g\colon Y\to Y'$, a 2-morphism is a Petri net morphism $\alpha\colon P\to P'$ making the following diagram commute, i.e.\ $\alpha\circ i=i'\circ Lf$ and $\alpha\circ o=o'\circ Lg$:
\[
\begin{tikzcd}
LX \arrow[r,"i"] \arrow[d,"Lf"'] & P \arrow[d,"\alpha"] & LY \arrow[l,"o"'] \arrow[d,"Lg"] \\
LX' \arrow[r,"i'"] & P' & LY' \arrow[l,"o'"']
\end{tikzcd}
\]
\end{enumerate}
A 2-morphism describes a map from one open Petri net to another that respects the boundaries, for example the inclusion of a smaller module into a larger one, or a map that merges redundant places and transitions.

\paragraph{Vertical composition.} Composition of vertical 1-morphisms is composition of functions. Two 2-morphisms that share a horizontal 1-cell,
\[
\begin{tikzcd}[column sep=small,row sep=small]
LX_1 \arrow[r,"i_1"] \arrow[d,"Lf"'] & P_1 \arrow[d,"\alpha"] & LY_1 \arrow[l,"o_1"'] \arrow[d,"Lg"] \\
LX_2 \arrow[r,"i_2"] & P_2 & LY_2 \arrow[l,"o_2"']
\end{tikzcd}
\qquad\text{and}\qquad
\begin{tikzcd}[column sep=small,row sep=small]
LX_2 \arrow[r,"i_2"] \arrow[d,"Lf'"'] & P_2 \arrow[d,"\beta"] & LY_2 \arrow[l,"o_2"'] \arrow[d,"Lg'"] \\
LX_3 \arrow[r,"i_3"] & P_3 & LY_3 \arrow[l,"o_3"']
\end{tikzcd}
\]
(note that the bottom of the first is the top of the second), compose vertically to
\[
\begin{tikzcd}[column sep=normal,row sep=small]
LX_1 \arrow[r,"i_1"] \arrow[d,"L(f'\circ f)"'] & P_1 \arrow[d,"\beta\circ\alpha"] & LY_1 \arrow[l,"o_1"'] \arrow[d,"L(g'\circ g)"] \\
LX_3 \arrow[r,"i_3"] & P_3 & LY_3 \arrow[l,"o_3"']
\end{tikzcd}
\]

\paragraph{Horizontal composition.} Two 2-morphisms that share a vertical 1-morphism $g$,
\[
\begin{tikzcd}[column sep=small,row sep=small]
LX \arrow[r,"i_P"] \arrow[d,"Lf"'] & P \arrow[d,"\alpha"] & LY \arrow[l,"o_P"'] \arrow[d,"Lg"] \\
LX' \arrow[r,"i_{P'}"] & P' & LY' \arrow[l,"o_{P'}"']
\end{tikzcd}
\qquad\text{and}\qquad
\begin{tikzcd}[column sep=small,row sep=small]
LY \arrow[r,"i_Q"] \arrow[d,"Lg"'] & Q \arrow[d,"\beta"] & LZ \arrow[l,"o_Q"'] \arrow[d,"Lh"] \\
LY' \arrow[r,"i_{Q'}"] & Q' & LZ' \arrow[l,"o_{Q'}"']
\end{tikzcd}
\]
compose horizontally to
\[
\begin{tikzcd}[column sep=large,row sep=small]
LX \arrow[r,"j_P\circ i_P"] \arrow[d,"Lf"'] & P+_{LY}Q \arrow[d,"\alpha+_{Lg}\beta"] & LZ \arrow[l,"j_Q\circ o_Q"'] \arrow[d,"Lh"] \\
LX' \arrow[r,"j_{P'}\circ i_{P'}"] & P'+_{LY'}Q' & LZ' \arrow[l,"j_{Q'}\circ o_{Q'}"']
\end{tikzcd}
\]
where $j_P,j_Q$ (and $j_{P'},j_{Q'}$) are the canonical maps into the pushouts, and $\alpha+_{Lg}\beta$ is the morphism induced by the universal property of the pushout $P+_{LY}Q$ \citep[Thm.~13]{BaezMaster2020}.

\paragraph{Tensor product.} The symmetric monoidal structure is given by disjoint union (coproducts in $\Set$ and $\Petri$). For 2-morphisms
\[
\begin{tikzcd}[column sep=small,row sep=small]
LX_1 \arrow[r,"i_1"] \arrow[d,"Lf_1"'] & P_1 \arrow[d,"\alpha_1"] & LY_1 \arrow[l,"o_1"'] \arrow[d,"Lg_1"] \\
LX'_1 \arrow[r,"i'_1"] & P'_1 & LY'_1 \arrow[l,"o'_1"']
\end{tikzcd}
\qquad\text{and}\qquad
\begin{tikzcd}[column sep=small,row sep=small]
LX_2 \arrow[r,"i_2"] \arrow[d,"Lf_2"'] & P_2 \arrow[d,"\alpha_2"] & LY_2 \arrow[l,"o_2"'] \arrow[d,"Lg_2"] \\
LX'_2 \arrow[r,"i'_2"] & P'_2 & LY'_2 \arrow[l,"o'_2"']
\end{tikzcd}
\]
the tensor product is
\[
\begin{tikzcd}[column sep=large,row sep=small]
L(X_1+X_2) \arrow[r,"i_1+i_2"] \arrow[d,"L(f_1+f_2)"'] & P_1+P_2 \arrow[d,"\alpha_1+\alpha_2"] & L(Y_1+Y_2) \arrow[l,"o_1+o_2"'] \arrow[d,"L(g_1+g_2)"] \\
L(X'_1+X'_2) \arrow[r,"i'_1+i'_2"] & P'_1+P'_2 & L(Y'_1+Y'_2) \arrow[l,"o'_1+o'_2"']
\end{tikzcd}
\]
using that $L$ preserves coproducts, $L(X_1+X_2)\cong LX_1+LX_2$. On horizontal 1-cells, the tensor product of two open Petri nets is their disjoint union, which describes the two nets running side by side without interacting; this is the open counterpart of the disjoint union $N_1+N_2$ in Example~\ref{ex:parallel}.

\subsection{Semantics of open Petri nets}\label{sec:semantics}
An open Petri net is a piece of syntax. Its meaning is given by \emph{semantics}: structure-preserving maps out of $\DOpenPetri$. Three are relevant here.

\paragraph{Operational semantics.} The functor $F\colon\Petri\to\CMC$ of Section~\ref{sec:monoids} extends to a symmetric monoidal double functor $\DOpenPetri\to\DOpenCMC$, sending an open Petri net $LX\to P\leftarrow LY$ to the open commutative monoidal category $FLX\to FP\leftarrow FLY$ \citep[Thm.~17]{BaezMaster2020}. This double functor preserves composition up to isomorphism: $F(Q\odot P)\cong FQ\odot FP$. Thus the processes of a composite net, in the sense of Section~\ref{sec:monoids}, can be computed from those of its parts. This is the precise sense in which the Meseguer--Montanari algebra of processes and open Petri nets fit together: the former is the operational semantics of the latter, not an alternative to it.

\paragraph{Reachability semantics.} For $P\colon X\pto Y$, the \emph{reachability relation} is
\[
\blacksquare P=\{(x,y)\in\Nset{X}\times\Nset{Y}\;:\;\Nset{o}(y)\text{ is reachable from }\Nset{i}(x)\text{ in }P\}.
\]
It records which markings of the outputs can be reached from which markings of the inputs, leaving no tokens elsewhere. \citet[Thms.~23 and 24]{BaezMaster2020} show that $\blacksquare$ is a \emph{lax} symmetric monoidal double functor into the double category of relations, which means that
\[
\blacksquare Q\circ\blacksquare P\subseteq\blacksquare(Q\odot P),
\]
and the inclusion can be strict: in the composite, tokens may pass from $P$ to $Q$ and back again, which the composite relation $\blacksquare Q\circ\blacksquare P$ does not see. Reachability of a composite can therefore be approximated from below, but not in general computed, from the reachability of its parts. \citet{BaezMaster2020} conjecture that equality holds for ``one-way'' open nets, in which no input place is in the target of a transition and no output place is in the source of a transition.

\paragraph{Rate-equation semantics.} \citet{BaezPollard2017} attach a positive rate constant $r(\tau)$ to each transition, which gives an \emph{open reaction network with rates}, and define a symmetric monoidal category $\RxNet$ of these (using decorated cospans of finite sets, with isomorphism classes as morphisms). Given $P\colon X\pto Y$ with places $S$, the law of mass action gives a vector field on $\R^S$,
\begin{equation}\label{eq:mass-action}
v(c)=\sum_{\tau\in T} r(\tau)\,\bigl(t(\tau)-s(\tau)\bigr)\prod_{\sigma\in S}c_\sigma^{\,s(\tau)_\sigma},
\end{equation}
where $s(\tau)_\sigma$ is the coefficient of $\sigma$ in $s(\tau)$, and the \emph{open rate equation}
\begin{equation}\label{eq:open-rate}
\frac{dc}{dt}=v(c)+i_*(I(t))-o_*(O(t)),
\end{equation}
where $I\colon\R\to\R^X$ and $O\colon\R\to\R^Y$ are arbitrary inflows and outflows and $i_*(I)_\sigma=\sum_{x:\,i(x)=\sigma}I_x$, $o_*(O)_\sigma=\sum_{y:\,o(y)=\sigma}O_y$. This assignment is a symmetric monoidal functor ``gray-boxing'' $\RxNet\to\Dynam$ into a category of open dynamical systems \citep{BaezPollard2017}. Concretely, the open rate equation of a composite $Q\odot P$ is obtained by identifying the outflows of $P$ with the inflows of $Q$ at each point of $Y$ and adding the right-hand sides. A further functor, ``black-boxing'', sends an open dynamical system to the relation between boundary concentrations and flows that holds in steady state \citep{BaezPollard2017}.

\subsection{Other notions of ``open'' Petri net}
The word ``open'' is used in other ways in the Petri net literature. \citet{Baldan2005} define an open net as a Petri net with two chosen subsets of places; up to replacing subsets by injections, these are the open Petri nets above whose input and output maps are injective, but \citet{Baldan2005} compose them differently and give a compositional semantics only for deterministic occurrence nets \citep[Sec.~4]{BaezMaster2020}. Nets with boundaries \citep{Rathke2014} compose along transitions rather than places. In this paper, ``open Petri net'' always means the notion of \citet{BaezMaster2020}.

\section{Open Petri nets for gene regulatory networks}\label{sec:grn-open}
We now apply the framework of Section~\ref{sec:open} to GRNs. The guiding idea is that a GRN module communicates with the rest of the network only through the genes it shares with it, so \emph{shared genes are the interface}.

\subsection{Modules and interfaces}\label{sec:rules}
A \emph{GRN module} is an open Petri net $P\colon X\pto Y$ built from the elementary modules of Section~\ref{sec:pn-grn}, where $X$ and $Y$ label the genes that $P$ shares with neighbouring modules. Section~\ref{sec:pushout} implies the following rules for decomposing a GRN into modules so that composition gives back the original network.
\begin{enumerate}[nosep,label=(R\arabic*)]
\item \emph{Cut through genes, not through interactions.} Composition glues places, never transitions. Every regulatory interaction (transition) must therefore lie entirely inside one module, and every gene that takes part in interactions in two modules must appear in the interface between them.
\item \emph{Assign each transition to exactly one module.} Since transitions are never identified, a transition placed in two modules appears twice in the composite and its effect is counted twice. This applies in particular to the degradation transition $d_g$ of a shared gene $g$: it belongs to exactly one of the modules that contain $g$.
\item \emph{Direction is bookkeeping.} The split of the interface into inputs and outputs fixes the order of composition only. Feedback between modules, for example a gene of a downstream module regulating a gene of an upstream module, needs no special treatment: both genes are simply included in the shared interface.
\end{enumerate}

\subsection{A worked example}\label{sec:example}
We decompose the GRN of Figure~\ref{fig:grn} into an upstream module, in which $A$ activates $B$ and $C$, and a downstream module, in which $B$ and $C$ regulate $D$ and $E$. The interface between the modules consists of the shared genes $B$ and $C$. We include a degradation transition for every gene (Remark~\ref{rem:limitations}), and assign $d_B$ and $d_C$ to the upstream module, in accordance with (R2).

\paragraph{Upstream module.} $P\colon X\pto Y$ has places $\{A,B,C\}$, transitions
\[
\T{B}{A}\colon A\to A+B,\quad \T{C}{A}\colon A\to A+C,\quad d_A\colon A\to 0,\quad d_B\colon B\to 0,\quad d_C\colon C\to 0,
\]
input set $X=\{x\}$ with $i(x)=A$ (so that $A$ can receive input, e.g.\ from an upstream signal), and output set $Y=\{y_1,y_2\}$ with $o(y_1)=B$, $o(y_2)=C$.

\paragraph{Downstream module.} $Q\colon Y\pto Z$ has places $\{B',C',D,E\}$, transitions
\[
\T{D}{B}\colon B'\to B'+D,\quad \T{D}{C}\colon C'+D\to C',\quad \T{E}{C}\colon C'\to E,\quad d_D\colon D\to 0,\quad d_E\colon E\to 0,
\]
input set $Y$ with $i(y_1)=B'$, $i(y_2)=C'$, and output set $Z=\{z_1,z_2\}$ with $o(z_1)=D$, $o(z_2)=E$. (We write $B',C'$ for the copies of $B,C$ in $Q$ to make the gluing visible.)

\paragraph{Composite.} By the construction of Section~\ref{sec:pushout}, the places of $Q\odot P$ are $\{A,B,C,D,E\}$, where $B$ and $C$ now denote the classes $\{B,B'\}$ and $\{C,C'\}$. Its transitions are the ten transitions of $P$ and $Q$, with sources and targets
\[
\begin{aligned}
&\T{B}{A}\colon A\to A+B, && \T{C}{A}\colon A\to A+C, && \T{D}{B}\colon B\to B+D, && \T{D}{C}\colon C+D\to C, && \T{E}{C}\colon C\to E,\\
&d_A\colon A\to 0, && d_B\colon B\to 0, && d_C\colon C\to 0, && d_D\colon D\to 0, && d_E\colon E\to 0 .
\end{aligned}
\]
These are exactly the transitions \eqref{eq:grn-transitions} of Figure~\ref{fig:grn}b, together with one degradation transition per gene. Hence $Q\odot P$ is the Petri net of Figure~\ref{fig:grn}b with degradation added (Figure~\ref{fig:grn-open}), now with an input at $A$ and outputs at $D$ and $E$ through which it can itself be composed with further modules. Because we work in $\DOpenPetri$ rather than with isomorphism classes, the places of the composite can still be identified by gene name.

\begin{figure}[htbp]
\centering
\begin{tikzpicture}[scale=0.85, every node/.append style={transform shape},
  bp/.style={pt}, bl/.style={font=\scriptsize}]
  \newcommand{\Pmodule}{%
    \node[place] (A) at (1.4,0.1) {$A$};
    \node[degr,label=below:{\scriptsize$d_A$}] (dA) at (1.4,-1.2) {};
    \node[vtrans,label=above:{\scriptsize$\T{C}{A}$}] (tCA) at (2.8,1.2) {};
    \node[vtrans,label=below:{\scriptsize$\T{B}{A}$}] (tBA) at (2.8,-1.0) {};
    \node[place] (C) at (4.2,1.2) {$C$};
    \node[place] (B) at (4.2,-1.0) {$B$};
    \node[degr,label=left:{\scriptsize$d_C$}] (dC) at (4.2,2.35) {};
    \node[degr,label=left:{\scriptsize$d_B$}] (dB) at (4.2,-2.15) {};
    \draw[arc] (A) -- (dA);
    \draw[rarc] (A) -- (tCA); \draw[arc] (tCA) -- (C);
    \draw[rarc] (A) -- (tBA); \draw[arc] (tBA) -- (B);
    \draw[arc] (C) -- (dC); \draw[arc] (B) -- (dB);}
  \newcommand{\Qmodule}[3]{%
    \node[vtrans,label=above:{\scriptsize$\T{E}{C}$}] (tEC) at (#3+1.4,1.2) {};
    \node[htrans,label=right:{\scriptsize$\T{D}{C}$}] (tDC) at (#3+2.1,0.1) {};
    \node[vtrans,label=below:{\scriptsize$\T{D}{B}$}] (tDB) at (#3+1.4,-1.0) {};
    \node[place] (E) at (#3+2.8,1.2) {$E$};
    \node[place] (D) at (#3+2.8,-1.0) {$D$};
    \node[degr,label=left:{\scriptsize$d_E$}] (dE) at (#3+2.8,2.35) {};
    \node[degr,label=left:{\scriptsize$d_D$}] (dD) at (#3+2.8,-2.15) {};
    \draw[arc] (#1) -- (tEC); \draw[arc] (tEC) -- (E);
    \draw[rarc] (#1) -- (tDC); \draw[arc] (D) -- (tDC);
    \draw[rarc] (#2) -- (tDB); \draw[arc] (tDB) -- (D);
    \draw[arc] (E) -- (dE); \draw[arc] (D) -- (dD);}
  \node[font=\small] at (-1.0,2.6) {(a)};
  \node[bnd,minimum height=5.0cm,label=above:$X$] at (0,0) {};
  \node[bp,label={[bl,label distance=2.4mm]left:$x$}] (x) at (0,0.1) {};
  \Pmodule
  \draw[leg] (x) -- (A);
  \node[bnd,minimum height=5.0cm,label=above:$Y$] at (5.6,0) {};
  \node[bp,label={[bl,label distance=2.4mm]right:$y_2$}] (y2) at (5.6,1.2) {};
  \node[bp,label={[bl,label distance=2.4mm]right:$y_1$}] (y1) at (5.6,-1.0) {};
  \draw[leg] (y2) -- (C); \draw[leg] (y1) -- (B);
  \node[font=\small] at (2.8,-3.0) {$P\colon X\pto Y$};
  \begin{scope}[xshift=7.4cm]
    \node[bnd,minimum height=5.0cm,label=above:$Y$] at (0,0) {};
    \node[bp,label={[bl,label distance=2.4mm]left:$y_2$}] (w2) at (0,1.2) {};
    \node[bp,label={[bl,label distance=2.4mm]left:$y_1$}] (w1) at (0,-1.0) {};
    \node[place] (Cp) at (1.4,1.2) {$C'$};
    \node[place] (Bp) at (1.4,-1.0) {$B'$};
    \draw[leg] (w2) -- (Cp); \draw[leg] (w1) -- (Bp);
    \Qmodule{Cp}{Bp}{1.4}
    \node[bnd,minimum height=5.0cm,label=above:$Z$] at (5.6,0) {};
    \node[bp,label={[bl,label distance=2.4mm]right:$z_2$}] (z2) at (5.6,1.2) {};
    \node[bp,label={[bl,label distance=2.4mm]right:$z_1$}] (z1) at (5.6,-1.0) {};
    \draw[leg] (z2) -- (E); \draw[leg] (z1) -- (D);
    \node[font=\small] at (2.8,-3.0) {$Q\colon Y\pto Z$};
  \end{scope}
  \begin{scope}[yshift=-6.6cm, xshift=2.2cm]
    \node[font=\small] at (-1.0,2.6) {(b)};
    \node[bnd,minimum height=5.0cm,label=above:$X$] at (0,0) {};
    \node[bp,label={[bl,label distance=2.4mm]left:$x$}] (x) at (0,0.1) {};
    \Pmodule
    \draw[leg] (x) -- (A);
    \Qmodule{C}{B}{4.2}
    \node[bnd,minimum height=5.0cm,label=above:$Z$] at (8.4,0) {};
    \node[bp,label={[bl,label distance=2.4mm]right:$z_2$}] (z2) at (8.4,1.2) {};
    \node[bp,label={[bl,label distance=2.4mm]right:$z_1$}] (z1) at (8.4,-1.0) {};
    \draw[leg] (z2) -- (E); \draw[leg] (z1) -- (D);
    \node[font=\small] at (4.2,-3.0) {$Q\odot P\colon X\pto Z$};
  \end{scope}
\end{tikzpicture}
\caption{The GRN of Figure~\ref{fig:grn} as a composite of two open Petri nets. (a) The upstream module $P$ and the downstream module $Q$, which share the genes $B$ and $C$ through the interface $Y$. Small rectangles with no outgoing arc are degradation transitions; those of the shared genes $B$ and $C$ belong to $P$ only. (b) The composite $Q\odot P$, obtained by identifying $B$ with $B'$ and $C$ with $C'$; it is the net of Figure~\ref{fig:grn}b with degradation added.}\label{fig:grn-open}
\end{figure}

\subsection{Compositional dynamics}\label{sec:grn-dynamics}
Assign rate constants $k_{BA},k_{CA},k_{DB},k_{DC},k_{EC}$ to the regulatory transitions and $\delta_A,\dots,\delta_E$ to the degradation transitions, and write the concentration of each gene product with the name of the gene. By \eqref{eq:mass-action} and \eqref{eq:open-rate}, the open rate equation of the upstream module $P$, with inflow $I_x$ at $A$ and outflows $O_{y_1},O_{y_2}$ at $B,C$, is
\begin{equation}\label{eq:rate-P}
\frac{dA}{dt}=-\delta_A A+I_x,\qquad
\frac{dB}{dt}=k_{BA}A-\delta_B B-O_{y_1},\qquad
\frac{dC}{dt}=k_{CA}A-\delta_C C-O_{y_2},
\end{equation}
and that of the downstream module $Q$, with inflows $I_{y_1},I_{y_2}$ at $B',C'$ and outflows $O_{z_1},O_{z_2}$ at $D,E$, is
\begin{equation}\label{eq:rate-Q}
\begin{aligned}
\frac{dB'}{dt}&=I_{y_1}, &
\frac{dC'}{dt}&=-k_{EC}C'+I_{y_2},\\
\frac{dD}{dt}&=k_{DB}B'-k_{DC}C'D-\delta_D D-O_{z_1}, &
\frac{dE}{dt}&=k_{EC}C'-\delta_E E-O_{z_2}.
\end{aligned}
\end{equation}
Note that $B'$ appears in $Q$ only as a regulator (read arc), so $Q$ by itself does not change it. To compose, we identify $B'=B$ and $C'=C$, set $I_{y_j}=O_{y_j}$, and add the right-hand sides of \eqref{eq:rate-P} and \eqref{eq:rate-Q} on shared genes. The flows across $Y$ cancel and we obtain
\begin{equation}\label{eq:rate-composite}
\begin{aligned}
\frac{dA}{dt}&=-\delta_A A+I_x, &\qquad
\frac{dB}{dt}&=k_{BA}A-\delta_B B,\\
\frac{dC}{dt}&=k_{CA}A-(k_{EC}+\delta_C)\,C, &
\frac{dE}{dt}&=k_{EC}C-\delta_E E-O_{z_2},\\
\frac{dD}{dt}&=k_{DB}B-k_{DC}CD-\delta_D D-O_{z_1}, &&
\end{aligned}
\end{equation}
which is exactly the mass-action rate equation of the composite net $Q\odot P$, as guaranteed by the functoriality of gray-boxing \citep{BaezPollard2017}. Equation~\eqref{eq:rate-composite} is a small ODE model of the GRN of the kind discussed in Section~\ref{sec:grn}: each gene has linear production driven by its activators and linear decay, and the inhibition of $D$ by $C$ appears as a $C$-dependent loss term.

For a constant input $I_x=I$ and no outflow at $Z$, the steady state is
\[
A^*=\frac{I}{\delta_A},\quad B^*=\frac{k_{BA}A^*}{\delta_B},\quad C^*=\frac{k_{CA}A^*}{k_{EC}+\delta_C},\quad
D^*=\frac{k_{DB}B^*}{\delta_D+k_{DC}C^*},\quad E^*=\frac{k_{EC}C^*}{\delta_E}.
\]
The expression for $D^*$ shows the repression of $D$ by $C$ as a hyperbolic (Hill coefficient one) dependence on $C^*$. Without degradation ($\delta_g=0$) no such steady state exists, which is point (i) of Remark~\ref{rem:limitations} seen dynamically. The relation that this steady state imposes between boundary concentrations and flows is what the black-boxing functor of \citet{BaezPollard2017} computes, and it too can be obtained by composing the steady-state relations of $P$ and $Q$.

\subsection{Qualitative behaviour and reachability}
Under the operational semantics, the processes of $Q\odot P$ are obtained by gluing those of $P$ and $Q$ \citep[Thm.~17]{BaezMaster2020}, so qualitative questions phrased in terms of processes are compositional. Reachability is subtler. Neither module above is one-way in the sense of Section~\ref{sec:semantics}: the input place $A$ of $P$ lies in the target of $\T{B}{A}$ and $\T{C}{A}$ (a read arc puts the regulator in both source and target), and the output place $D$ of $Q$ lies in the source of $\T{D}{C}$ and $d_D$. More generally, non-consuming regulation and degradation make GRN modules rarely one-way. For such modules the composite of the reachability relations of the parts is only guaranteed to be contained in the reachability relation of the whole \citep{BaezMaster2020}. Compositional reachability analysis of GRNs therefore yields under-approximations, and exact reachability questions should be asked of the composite net.

\section{Discussion and future work}\label{sec:conclusion}
We have shown how open Petri nets \citep{BaezMaster2020} can serve as a compositional language for GRNs. Treating shared genes as interfaces, a GRN can be cut into modules, each an open Petri net, and recovered exactly by composition, provided interactions are not split across modules and each transition is assigned to one module (Section~\ref{sec:rules}). The double category $\DOpenPetri$ keeps track of which place is which gene in the composite, and maps between open Petri nets (2-morphisms) give a precise way to say that one module is a refinement or simplification of another. Different semantics then behave differently under composition: the operational semantics and the mass-action rate equation \citep{BaezPollard2017} are compositional, whereas reachability is only laxly so.

The approach has several limitations. The encoding of Section~\ref{sec:pn-grn} is deliberately simple, with abstract tokens and linear mass-action kinetics, whereas many GRN models use Hill-type or Boolean regulation functions and inhibitor arcs, which lie outside the class of nets considered here (Remark~\ref{rem:limitations}). Mass-action models with explicit promoter or complex-formation steps can produce Hill-like responses while remaining within the framework, at the cost of more places and transitions. Composition along places also requires that the modules agree on how each shared gene is represented.

Several directions remain for future work: deriving open Petri net modules from GRNs inferred from expression data, for example single-cell RNA sequencing; extending the construction to coloured, stochastic or fuzzy Petri nets, which are widely used in systems biology \citep{LiuHeinerGilbert2019,LiuHeinerGilbert2020} and which \citet{BaezMaster2020} expect to admit similar treatments; and using software that implements open Petri nets, such as AlgebraicPetri.jl \citep{AlgebraicPetri}, to build and analyse large GRNs from libraries of reusable modules.

\section*{Acknowledgements}
The author thanks Claude Opus 5.5 (Anthropic) for assistance in revising this manuscript, including checking it against the cited literature, correcting the categorical constructions and preparing the figures.

\bibliographystyle{apalike}
\bibliography{OpenPetriNet_GRN_v2}

\end{document}